\documentclass[journal=ancac3,manuscript=article,layout=standard]{achemso}

\usepackage[version=3]{mhchem} 
\usepackage{amsmath}
\usepackage{amssymb}
\usepackage{textcomp}
\usepackage{graphicx}
\usepackage[dvipsnames]{xcolor}
\usepackage{float}

\author{Justine Baronnier}
\affiliation{Institut Lumi{\`e}re-Mati{\`e}re, CNRS UMR5306, Universit{\'e}
Lyon 1, Universit{\'e} de Lyon, 69622 Villeurbanne CEDEX, France}

\author{Julien Houel}
\affiliation{Institut Lumi{\`e}re-Mati{\`e}re, CNRS UMR5306, Universit{\'e}
Lyon 1, Universit{\'e} de Lyon, 69622 Villeurbanne CEDEX, France}
\email{julien.houel@univ-lyon1.fr}

\author{Christophe Dujardin}
\affiliation{Institut Lumi{\`e}re-Mati{\`e}re, CNRS UMR5306, Universit{\'e}
Lyon 1, Universit{\'e} de Lyon, 69622 Villeurbanne CEDEX, France}

\author{Florian Kulzer}
\affiliation{Institut Lumi{\`e}re-Mati{\`e}re, CNRS UMR5306, Universit{\'e}
Lyon 1, Universit{\'e} de Lyon, 69622 Villeurbanne CEDEX, France}

\author{Benoît Mahler}
\affiliation{Institut Lumi{\`e}re-Mati{\`e}re, CNRS UMR5306, Universit{\'e}
Lyon 1, Universit{\'e} de Lyon, 69622 Villeurbanne CEDEX, France}
\email{benoit.mahler@univ-lyon1.fr}

\title{Doping \ce{MAPbBr3} hybrid perovskites with CdSe/CdZnS quantum dots: from emissive thin-films to hybrid single-photon sources}


\begin{document}




\clearpage

\begin{abstract}

\noindent We report the first doping of crystalline methyl\-ammonium lead bromide (\ce{MAPbBr3}) perovskite thin-films with CdSe/CdZnS core/shell quantum dots (QDs), using a soft-chemistry method that preserves their high quantum yield and other remarkable fluorescence properties. Our approach produces \ce{MAPbBr3} films of 100\,nm thickness doped at volume ratios between 0.025 and 5\,\% with colloidal CdSe/CdZnS QDs whose organic ligands are exchanged with halide ions to allow for close contact between the QDs and the perovskite matrix. Ensemble photoluminescence (PL) measurements demonstrate the retained emission of the QDs after incorporation into the \ce{MAPbBr3} matrix. Ensemble photoluminescence excitation (PLE) spectra exhibit signatures of wavelength-dependent coupling between the CdSe/CdZnS QDs and the \ce{MAPbBr3} matrix, i.\,e., a transfer of excitation energy from matrix to QD or from QD to matrix. Spatially-resolved PL experiments reveal a strong correlation between the positions of QDs and an enhancement of the PL signal of the matrix. Fluorescence lifetime imaging (FLIM) of the doped films furthermore show that the emission lifetime of \ce{MAPbBr3} is slower in the vicinity of QDs, which, in combination with the increased PL signal of the matrix, suggests that QDs can act as local nucleation seeds that improve the crystallinity of \ce{MAPbBr3}, thus boosting its emission quantum yield. Confocal PL-antibunching measurements provide clear evidence of single-photon emission from individual QDs in perovskite. Finally, the analysis of blinking statistics indicates an improvement of the photostability of individual QDs in perovskite as compared to bare CdSe/CdZnS QDs. At high CdSe/CdZnS QD doping levels, this work opens thus the route to hybrid solar concentrators for visible-light harvesting and hybrid-based LEDs. Finally, low-doping content would lead to hybrid single-photon sources embedded in field-effect devices for single charge control to serve as an alternative to solid-state quantum dots and open the route to build nanophotonic devices with high-quantum-yield CdSe-based colloidal QDs.

\end{abstract}


\section{INTRODUCTION}

Research on hybrid perovskites for photovoltaics has been booming since 2009, evolving from the first reported efficiency of 3.8\,\% for a solar cell using perovskites as photo-active materials \cite{kojima09may6} to an efficiency of 28\,$\%$ in 2019\cite{nayak19natre4269}. Since then, the literature and potential applications of hybrid perovskites have grown exponentially\cite{wei19mar6, akkerman18natre5394, fu19natre3169, mykhaylyk19oct1}. In addition to their remarkable optical and transport properties, this success of hybrid halide perovskites is explained by them being easy to process and allowing to create multi-scale compounds (from bulk\cite{yu18dec17} to thin-films\cite{stranks15natna5391} and quantum dots (QDs) \cite{utzat19mar8}), as well as by their compatibility with industrially-scale soft-chemistry processes \cite{fu19natre3169,park16oct17}. However, creating heterostructures with hybrid perovskites as active materials has so far remained an elusive goal. An important step to address this issue was recently taken with the use of hybrid perovskites (methylammonium lead iodide, MAPI) to create a host matrix for a different active material (PbS QDs) \cite{ning15jul16}. However, this first QD-in-perovskite hybrid offers only limited perspectives for luminescence applications, on the one hand because PbS NCs emit in the infrared, which is not ideal for solar concentrators or LED applications and on the other hand because their low fluorescence quantum yield (QY) has so far not allowed the observation of single nano-objects, which would be essential for potential applications as single-photon sources (SPS). Following that first milestone, several promising ways to taking advantage of the properties of II-VI QDs and hybrid perovskites have been proposed: A thin film containing CdSe and \ce{CsPbI3} QDs has been deposited on top of \ce{MAPbI3} to study potential increases in conversion efficiency of perovskite-based solar cells \cite{Ge20}. PbSe nanowires have been coated with \ce{CsPbBr3} for applications as photodetectors \cite{Fan18}. \ce{CsPbBr3}/CdS core/shell QDs with enhanced stability have been synthesized \cite{Tang19}. Another recent example is the creation of \ce{CsPbBr3}/ZnS heterodimers and heterostructures to enhance stability \cite{Ravi20}. In spite of these important developments, no successful realization of CdSe-based QD-in-perovskite hybrid for luminescence applications has been reported so far to the best of our knowledge. CdSe colloidal quantum dots (QDs) are an economical, easy-to-tune, highly fluorescent active material \cite{bawendi_synthesis_1995, mahler08natma7659}. The chemical synthesis of QDs is already established at the industrial scale and can be found in consumer products. Additional advantages of QDs are their tunable emission with nanosecond radiative recombination lifetimes and their tunable sheet density, which can be orders of magnitude higher than what is achievable by epitaxial growth. As a consequence, substantial research effort has been dedicated to CdSe QDs in LEDs\cite{bozyigit13jun25}, solar cells\cite{zhang12chemc9111235,huang16dec21}, solar concentrators\cite{hyldahl09solen4566} or as SPS and building blocks for quantum-information devices \cite{lounis00oct27, biadala09jul17, fernee12natco31287, fernee13nov22}. Unfortunately, the fragility of this class of emitters during  encapsulation process \cite{woggon05nanol5483, rashad10physs71523} and the lack of compatible semiconductor host matrices has rendered difficult their incorporation in crystalline, fast charge-transfer devices analog to what was achieved with state-of-the-art InGaAs diodes \cite{warburton00jun22}. The semiconductor nature of hybrid perovskites and the process compatibility between CdSe-QD synthesis and perovskite thin-film elaboration (soft chemistry) suggests that combining the remarkable properties of these two materials is an intrinsically interesting route for applications such as light converters (solar concentrator or LEDs at high doping level) and single photon sources (at low doping level). For the last two years, an intense effort has been put on coupling high quantum yield CdSe-based QDs with either organic or inorganic hybrid perovskite \cite{wei19mar6, akkerman18natre5394, fu19natre3169, ge20apr1, li19oct30}. So far, research has reported on either depositing CdSe QDs over a perovskite layer to enhance its light harvesting properties \cite{li19oct30}, or insertion of CdSe and \ce{CsPbI3} in between a \ce{MAPbI3} layer and a hole transport layer \cite{ge20apr1}. However, to the best of our knowledge, none of the results reported so far was able to retain the fluorescence efficiency and single photon emission capability, which are actually the best assets of CdSe-based colloidal QDs.


In this article, we present the first emissive QD-in-perovskite hybrids in the visible range, based on thin \ce{MAPbBr3} films doped with highly fluorescencent CdSe/CdZnS core/shell quantum dots. We present ensemble photoluminescence (PL) spectra demonstrating emission of these films at the CdSe/CdZnS wavelength. Photoluminescence excitation (PLE) experiments were performed, which demonstrate different types of response and coupling between the CdSe/CdZnS QDs and the \ce{MAPbBr3} matrix. The combination of confocal PL microscopy and fluorescence lifetime imaging (FLIM) reveal sub-micron-scale correlations between the emission of the QDs and the \ce{MAPbBr3} matrix. Finally, confocal PL microscopy demonstrates the first QD-in-perovskite hybrid single photon source emitting in the visible, with blinking properties enhanced compared to the non-encapsulated QDs.


\section{METHODS}

\subsection{Quantum Dot Synthesis}

 CdSe/Zn$ _{1-x}$Cd$_x$S Colloidal quantum dots (CQDs) were synthesized following Refs.~\citenum{lim14dec17} and \citenum{Baronnier21}. To create the CdSe core, 15\,mL of octadecene, 1\,mmol of cadmium oxide and 3\,mmol of  myristic acid were introduced in a three-neck flask. The mixture was degassed at 60\,$^{\circ}$C under vacuum for ten minutes before being heated to 300\,$^{\circ}$C under argon. When the mixture became transparent, 0.25\,mL of TOPSe at 2\,M was introduced to react at 300\,$^{\circ}$C for three minutes. The Zn$\rm _{x-1}Cd_x$S shell was then created: 3\,mL of zinc oleate (\ce{Zn(OA)2}) at 0.5M and 1\,mmol of dodecanthiol (DDT) were injected during one minute. The solution was then left to react 300$^{\circ}$C for 30 minutes. The shell created in this first growth step was composed theoretically of 60\,\% of Cadmium and 20\,\% of Zinc. Then, 2\,mL of cadmium oleate (\ce{Cd(OA)2}) at 0.5\,M, 4\,mL of \ce{Zn(OA)2} at 0.5\,M and 1.5\,mL of TOPS at 2\,M were injected in one minute, resulting in a final shell that was equal parts cadmium and zinc. This mixture was heated during 10 minutes to 300\,$^{\circ}$C. This step was repeated 4 times and for each injection, a few nanometers of shell were created. The synthesized QDs were eventually dispersed in hexane.

\subsection{Ligand Exchange}

To transfer the CQDs in the perovskite precursor solvent, which is N,N-Dimethylformamide (DMF), an OA-to-halide ligand exchange was necessary. To this end, 60\,\textmu L of  CQDs  were  injected  into  2\,mL of hexane containing 1\,mL of methylammonium bromide (0.2\,M) and cadmium  bromide (0.3\,mmol). This solution was stirred until the QDs were transferred from hexane to DMF, at which point they were precipitated with isopropanol and redispersed in DMF two times. In the end, CQDs with bromide ligands were dispersed in DMF (2\,mL) and 1\,mL of methyammonium chloride (0.1\,M) then the solution was sonicated for 25 minutes at 50\,$^{\circ}$C. An extensive optical characterization of the obtained halide-covered CdSe/CdZnS QDs can be found in Ref.~\citenum{Baronnier21}.

\subsection{QD-in-Perovskite Heterostructures} 

Thin-film QD-in-perovskite heterostructures were created by spin-coating a solution of perovskite precursor and halide-ligand QDs onto a glass (BK7) slide: 100\,\textmu L of perovskite precursor solution at 1\,M and 100\,\textmu L of QDs with the appropriate concentration were spin-coated at 5000\,rpm during 60 seconds; 5 seconds after the start, 1\,mL of chlorobenzene was injected. 

\subsection{Ensemble spectroscopy}

PL and PLE experiments were performed with an FS5 spectrofluorometer (Edinburgh Instr.) at 405\,nm excitation wavelength for the PL spectra, and 645\,nm detection for PLE measruements. Resolution for both experiments was set at 1\,nm, and integration time per point at 300\,ms and 1\,s for PL and PLE respectively.

\subsection{Spatially Resolved PL/FLIM Measurements}

Spatially resolved measurements were performed with the home-built confocal microscope described in Ref.~\citenum{Aubret_2020}.
Images were recorded with an oil-immersion objective (Olympus NA=1.35, x60) which focused a pulsed
laser-light (Edinburgh Instr. EPL0450) at 445\,nm and 2\,MHz repetition rate, and collected the emitted
signal. The intensity used was 50\,W/cm$^2$ for sample A, and 150\,W/cm$^2$ for sample B. The images
are presented as the sum of 10 consecutive scans with 100\,ms integration time per pixel per scan.
The images were corrected for the small in-plane drift of less than 2 pixels that occurred
during the measurements. Detection was assured by two single-photon avalanche photodiodes (SPADs),
an ARQH-050 (Perkin Elmer) for QD luminescence and a Count-20 C (Laser Components) for the \ce{MAPbBr3} signal; these detectors were connected to a time-correlated single-photon counting card
(TimeHarp 200 with router, PicoQuant). The Bandpass filters used for selective detection were a combination of an FBH0650 (Thorlabs) with a BrightLine 655 (Semrock) for QD emission and a 65-700
(Edmund Optics) with an OD=1 neutral density filter for the \ce{MAPbBr3} signal.

\subsection{Individual QDs in Perovskite}

Measurements were performed using the same home-built set-up,. Antibunching and blinking measurements were performed with a cw laser excitation at 561\,nm (Exelitas). A bandpass filter (Thorlabs FBH0650 and Semrock BrightLine 655) was placed in front of both SPADs to select the emission of individual QDs in the QD-in-perovskite hybrid for the antibunching measurements.

\section{RESULTS AND DISCUSSION}

\subsection{Undoped Perovskite Thin Films}

\begin{figure}[t]
\includegraphics[clip,width=15cm]{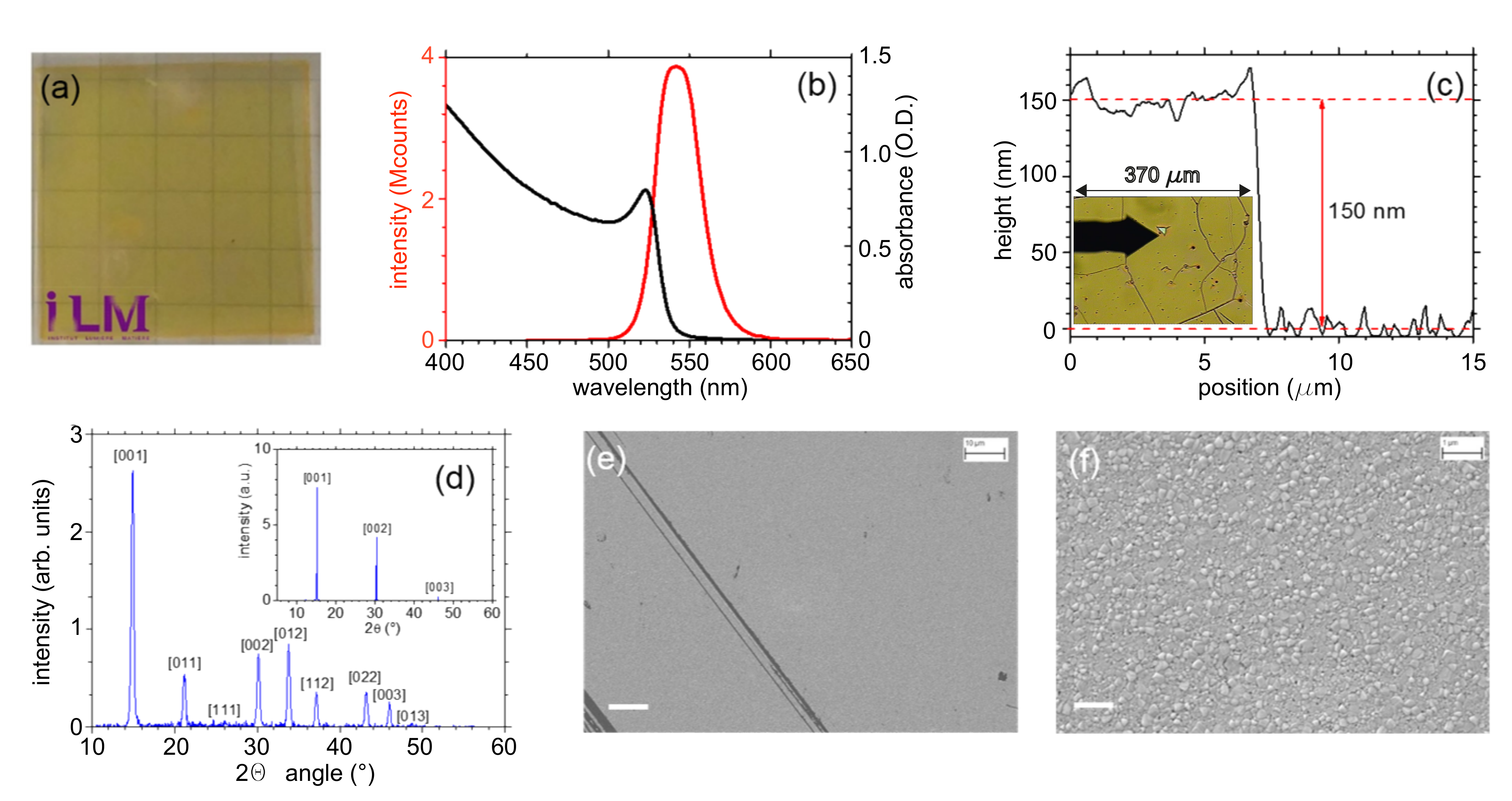}
\caption{Characterization of thin films of undoped \ce{MAPbBr3} deposited on glass. (a) White-light photograph. (b) Absorption spectrum (black) and PL emission spectrum for an excitation wavelength of 405\,nm (red). (c) AFM thickness measurement. (d) X-ray diffraction diagram under grazing incidence and under standard incidence (inset). (e) SEM image of a 110$\times$76\,\textmu m$^2$ area; the scale bar indicates 10\,\textmu m. (f) Zoomed-in SEM image (scale bar: 1\,\textmu m).}
\label{figure1}
\end{figure}

We present in Fig.~\ref{figure1} optical and structural characterizations of undoped \ce{MAPbBr3} films. Figure~\ref{figure1}\,(a) shows a photograph of a typical film obtained with the protocol detailed in Methods: transparent-orange, indicating absorption in the visible and low-diffusing behavior as a hint to a good crystallinity. The absorption spectrum of the same sample,  Fig.~\ref{figure1}\,(b), exhibits a strong and narrow absorption peak at 520\,nm typical of an \ce{MAPbBr3} exciton \cite{zhao18acsen71662}. The corresponding photoluminescence (PL) spectrum shows a narrow emission peak centered at 540\,nm, with a full width at half-maximum (FWHM) of 35\,nm. We note that the emission is actually structured by two peaks at $\lambda$ = 533 and 544\,nm, with intensity ratio varying over different region of the sample. The emission at 533\,nm is attributed to the band-edge free exciton \cite{Guo17, Liu18, kumar19jun6}, and the emission at 544\,nm to shallow defects near the \ce{MAPbBr3} band-edge \cite{Liu18, Motti19}. Surprisingly, the shallow defects do not have a signature in the \ce{MAPbBr3} absorption spectrum, but the unambiguous detection of their emission points to a vanishing absorption coefficient and their emission being activated \emph{via} charge relaxation from the \ce{MAPbBr3} band-edge exciton \cite{Liu18, kumar19jun6}. Atomic force microscopy (AFM) of the film reveals a thin-film thickness of 150\,nm, Fig.~\ref{figure1}\,(c). Fig.~\ref{figure1}\,(c) is an AFM profile taken from a crack on the sample surface presented in the optical microscope picture in the inset of Fig.~\ref{figure1}\,(c) (the dark region is the shadow of the AFM cantilever). Thickness measurement were performed on 8 samples and returned an average thickness of 152\,nm. This is the average thickness measured on undoped \ce{MAPbBr3} thin-films, created with 1\,M concentration of the precursor. The protocol to insert the QDs leads to a 0.5\,M concentration of the perovskite precursor and average doped film thickness of $\sim$100\,nm. Figure~\ref{figure1}\,(d) presents an X-ray diffraction diagram of the sample in Fig.~\ref{figure1}\,(a, b) realized under grazing incidence; peaks are labelled according to the crystallographic planes of \ce{MAPbBr3}. The inset in Fig.~\ref{figure1}\,(d) presents the diffraction diagram in standard geometry, in which only $[0\, 0 X]$ directions with $X=1$, 2 and 3 contribute, indicating a strong orientation of the thin film along those axes. The crystallization of the film thus occurs without any annealing step, but is induced by the injection of the chlorobenzene counter-solvent at 4\,s after the spin-coating started \cite{Yu17,Yang19}. We present in Fig.~\ref{figure1}\,(e) a scanning electron microscopy (SEM) image of the sample over a 110x76\,\textmu m$^2$ region, showing an homogeneous surface with occasional straight-line cracks, with an AFM-measured depth of 150\,nm (Fig.~\ref{figure1}\,(c)). A more resolved SEM picture over a region of 11x7.5\,\,\textmu m$^2$, Fig.~\ref{figure1}\,(f), shows the crystallite fine structure, with sub-micron size. The arrangement suggests for a compact thin film of \ce{MAPbBr3}.

\subsection{QD-in-perovskite hybrid: ensemble measurements}

\begin{figure}[t]
\includegraphics[clip,width=15cm]{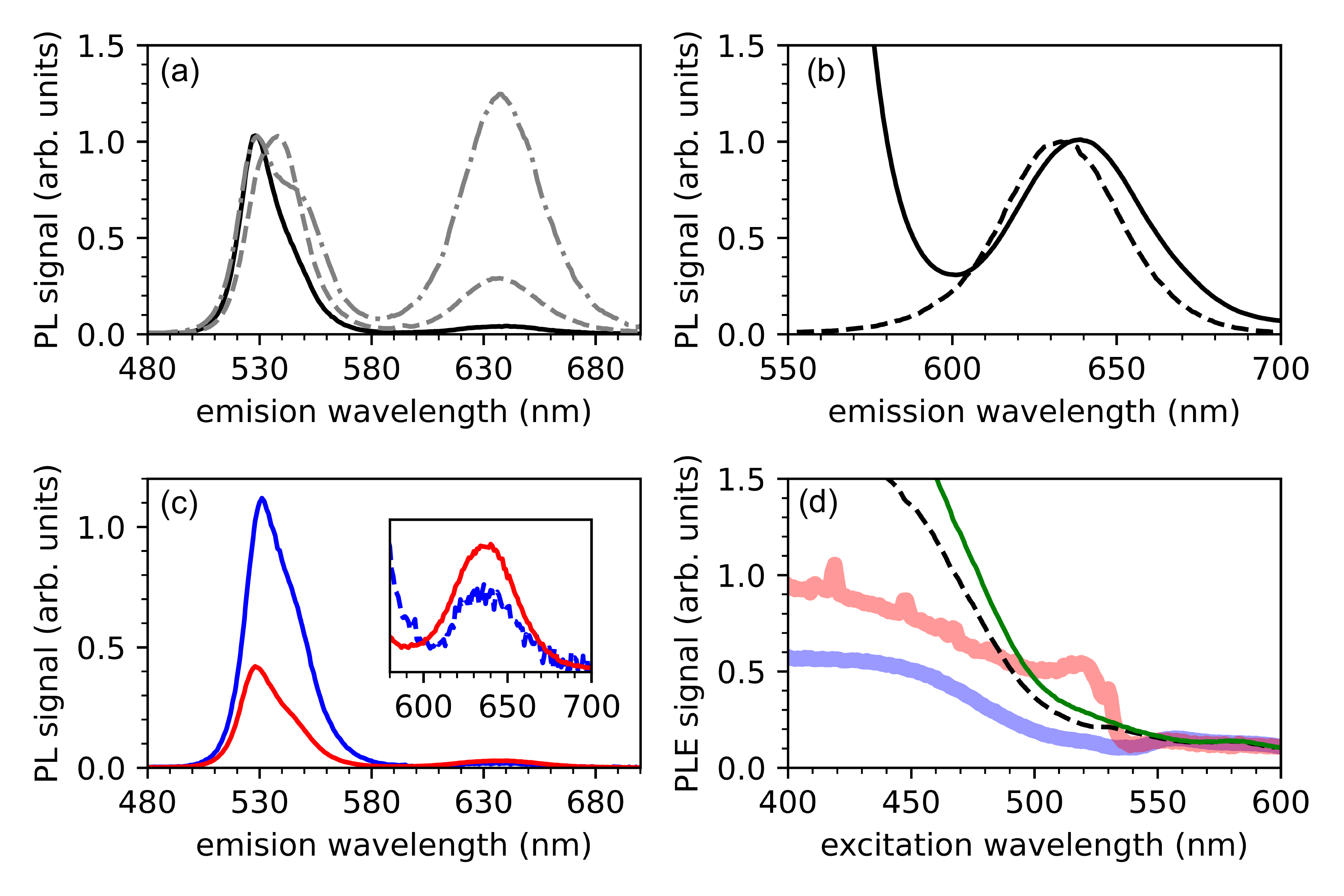}
\caption{(a) PL emission spectra of \ce{MAPbBr3} doped with CdSe/CdZnS QDs at three different levels: 0.1\,\% (solid line), 0.5\,\% (dashed line) and 5\,\% (dash-dotted line); excitation wavelength: 405\,nm. (b) Comparison between PL spectra of CdSe/CdZnS QDs on glass (dashed line) and for of QDs in perovskite film (solid line); excitation wavelength: 405\,nm. (c) PL spectra of sample A (nominal doping 0.025\,\%, blue) and B (nominal doping 0.1\,\%, red). (d) PLE spectra of sample A (blue) and B (red), together with a reference PLE of CdSe/CdZnS QDs on glass (green) and the expected PLE spectrum (dashed black) of CdSe/CdZnS QD in \ce{MAPbBr3} in the absence of any interaction with the matrix (see text).}
\label{figure2}
\end{figure}

Figure~\ref{figure2}\,(a) shows the photoluminescence (PL) spectra of three different \ce{MAPbBr3} thin films doped at 0.1, 0.5, and 5\,\% with CdSe QDs. The spectra are normalized to their respective perovskite maximum emission. The luminescence of the QDs can be observed even at a doping level of  0.1\,\%, centered at 638\,nm, with an emission linewidth of 43\,nm. Increasing the doping level to 0.5 and 5\,\% increases the QD luminescence with respect to the perovskite matrix emission, confirming that emission survives from low to high doping levels. The emission wavelength and the full width at half maximum (FWHM) remain unchanged at 638\,nm and 43\,nm respectively for all doping levels. On the other hand, the center emission wavelength of the QD-in-perovskites hybrid is red-shifted to the red by 4\,nm compared to QDs in air, Fig.\ref{figure2}\,(b). We attribute the shift to the heteroepitaxial relationship between the perovskite and the QDs created at the hybrid interface, in which the surrounding perovskite acts a shell to modify the potential barrier seen by the exciton. 


\begin{figure}[H]
\includegraphics[clip,width=15cm]{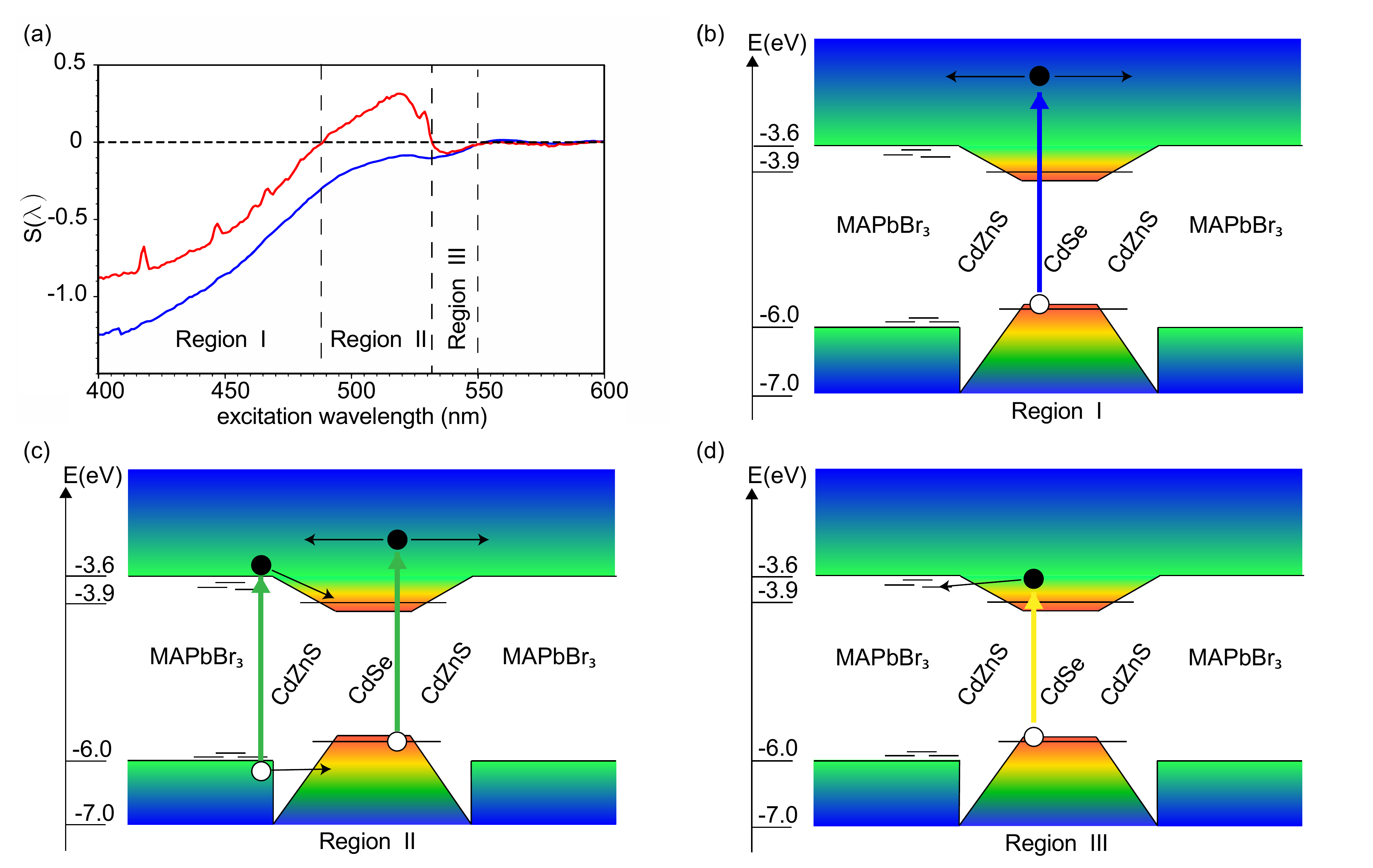}
\caption{(a): $S_\text{A,B}(\lambda)$ as defined in the text, shown as a function of the excitation wavelength for sample A (blue) and sample B (red). (b), (c) and (d): Proposed band diagrams for the energy levels of the QD-in-perovskite hybrids (energy levels for \ce{MAPbBr3} are taken as the mean values of refs\cite{ADJOKATSE2017413, Noh13, Seungchan14, Chen16, Zhang18, Schulz14, kim16}), with associated coupling processes for Region I (b), Region II (c) and Region III (c). Electrons and holes are depicted as black and white dots, respectively, the excitation as colored arrows (blue, green and yellow) and the coupling processes as black arrows. (b): Blue-light excitation promotes an electron to the conduction band of the QD-in-perovskite hybrid. The electron can diffuse in the \ce{MAPbBr3} matrix, which diminishes the probability of a photon emitted by the QDs. (c): Green-light excitation can promote an electron to the \ce{MAPbBr3} continuum from the QD valence band. The electron can diffuse in the perovskite, and the probability to relax in the QD is decreased compared to the case of QD on glass. Green light can also create a \ce{MAPbBr3} band-edge exciton in the immediate vicinity of a QD, followed by exciton transfer to the QD. These two process are competing with different weights for sample A and B. (d) Yellow light promotes an electron from the QD valence band to its conduction band. 
This electron can relax to nearby shallow defects of the \ce{MAPbBr3} matrix, which decreases the probability for the QD emitting a photon.}
\label{figure3}
\end{figure}

We present in Fig.\ref{figure2}\,(c) the PL spectra of two CdSe/CdZnS-doped thin films: Sample A (blue) and sample B (red), prepared under conditions corresponding to a nominal doping of 0.025 and 0.1\,\%, respectively. The two spectra are comparable, with strong \ce{MAPbBr3} emission around 530\,nm and a weaker QD contribution centered at $\lambda$ = 638\,nm (see inset in Fig.\ref{figure2}\,(c)). Photoluminescence excitation (PLE) spectra on sample A and B, detected at the QDs emission wavelength ($650 \pm 1$\,nm) are shown in Fig.~\ref{figure2}\,(d), with two reference spectra, QD on glass on the one hand and, on the other hand,
the expected PLE of a hypothetical interaction-free \ce{MAPbBr3} film doped homogeneously with QDs. The first noticeable feature of these results is the difference between the PLE obtained from sample A and sample B. The second highlight is the strong deviation of these two PLE spectra from what is expected from QDs in a \ce{MAPbBr3} matrix in the absence of coupling. To help quantify and explain these differences, we introduce the quantity $S(\lambda)$:
\begin{equation}
S_\text{A,B}(\lambda) = S^\text{(exp)}_\text{A,B}(\lambda) - S_0(\lambda)
\quad,
\end{equation}
where $S^\text{(exp)}_\text{A,B}(\lambda)$ is the experimental PLE spectrum of sample A and B, respectively. The expected PLE spectrum for QDs in \ce{MAPbBr3} without any interaction between them and the host matrix is $S_0(\lambda) = S_\text{ref}(\lambda) \bigl( 1-e^{-\alpha(\lambda) d} \bigr)$, where $S_\text{ref}(\lambda)$ represents the measured PLE spectrum of bare QDs on glass, while $\alpha(\lambda)$ and $d$ stand for the absorption coefficient and the thickness of the \ce{MAPbBr3} host matrix respectively. Given this definition of $S_\text{A,B}(\lambda)$, we expect $S_\text{A,B}(\lambda) = 0$ for samples in which the QDs do not interact with the host matrix, while $S_\text{A,B}(\lambda) \neq 0$ indicates an exchange of energy between the matrix and the QDs. As can be seen in Fig.~\ref{figure3}\,(a), there are indeed two main wavelength zones: one in which $S_\text{A,B}(\lambda) \simeq 0$ for $\lambda\gtrsim550$\,nm, which corresponds to the transparency region of the perovskite where we do not expect coupling between the QD and the \ce{MAPbBr3} matrix, and another one with $S_\text{A,B}(\lambda) \neq 0$ for $\lambda\lesssim550$\,nm, which is the absorbing region of the \ce{MAPbBr3} matrix. We can furthermore distinguish three regions in that latter zone, Region I ($400\,\text{nm} \lesssim \lambda \lesssim 490$\,nm), II ($490\,\text{nm} \lesssim \lambda \lesssim 530$\,nm) and III ( $530\,\text{nm} \lesssim \lambda \lesssim 550$\,nm). In Region I, we find $\rm S_{A,B}(\lambda) < 0$ for both sample A and B, meaning that coupling between the QD and the perovskite occurs such that it leads to a less efficient QD emission. We explain this signal-decrease in Fig.~\ref{figure3}\,(b), where we indicate the ancillary relaxation paths of an electron promoted to the conduction band of QDs in perovskite that are not available in the absence of the matrix. While the main relaxation path of the electron for the bare QDs leads to the QD conduction band edge, the hybrid sample makes it  becomes possible for the electron to be delocalized in the perovskite matrix. This effect gets stronger as the excitation energy is increased, since the probability to fall- back in the QD decreases. In Region II, the optical response of sample A and B are different: $S_\text{A}(\lambda) < 0$ while $S_\text{B}(\lambda) > 0$. QDs in sample A lose signal to the perovskite, by the same mechanism as in region I, while QDs in sample B gain signal from the perovskite. Figure~\ref{figure3}\,(c) proposes a scenario in which coupling between the QDs and the perovskite matrix can lead to an increase of the QD signal: The putative band alignment of our QD-in-perovskite system suggests that Region II could correspond to the creation of excitons at the \ce{MAPbBr3} band edge ($\lambda \sim 521$\,nm), and that these excitons can actually be transferred to a nearby QD by electron relaxation and hole tunneling.
This process could be conceptualized as the surrounding perovskite acting as a shell for the QD, increasing its effective absorption cross-section for wavelengths in Region II, leading to a competition between the effects described for Region I (Fig.~\ref{figure3}\,(b)) and Region II (Fig.~\ref{figure3}\,(c)), which is sample-dependent. We speculate that the local hetero-epitaxial arrangement between the QDs and the perovskite matrix might be different for sample A and B, leading to different contributions of the two coupling phenomena (from QD to matrix vs.\ from matrix to QD). Since the crystallization of the compound highly depends on the timing of the chlorobenzene injection, which is carried out manually, it is not surprising that small discrepancies from sample to sample can lead to differences in the QD-perovskite interface; further investigation will be necessary to achieve a deeper understanding of this phenomenon. In Region III, both sample A and B exhibit comparable behavior with $S_\text{A,B}(\lambda) \lesssim 0$. This is surprising at first since Region III is mostly transparent if one looks at the \ce{MAPbBr3} absorption spectrum (Fig.~\ref{figure1}\,(b)). However, the \ce{MAPbBr3} and sample A and B emission spectra, clearly exhibit a sub-band-edge contribution that we attribute to the emission of shallow defects \cite{kumar19jun6, delport19sep5}. These defects do not show a clear signature on the absorption spectra, implying a small absorption cross-section. However, their presence close to the \ce{MAPbBr3} band-edge could allow tunneling of an electron out of the QD to these defects states, lowering the QD emission signal when excited in Region III of the spectrum, as depicted in Fig.~\ref{figure3}\,(d). Confirming these explanations would require further studies, which should include variation of the \ce{MAPbBr3} matrix composition with \ce{Cl} and \ce{I} halides to vary the band structure. 


\subsection{QD-in-Perovskite Hybrid: Spatially Resolved Measurements}

\begin{figure}[t]
\includegraphics[clip,width=15cm,angle=0]{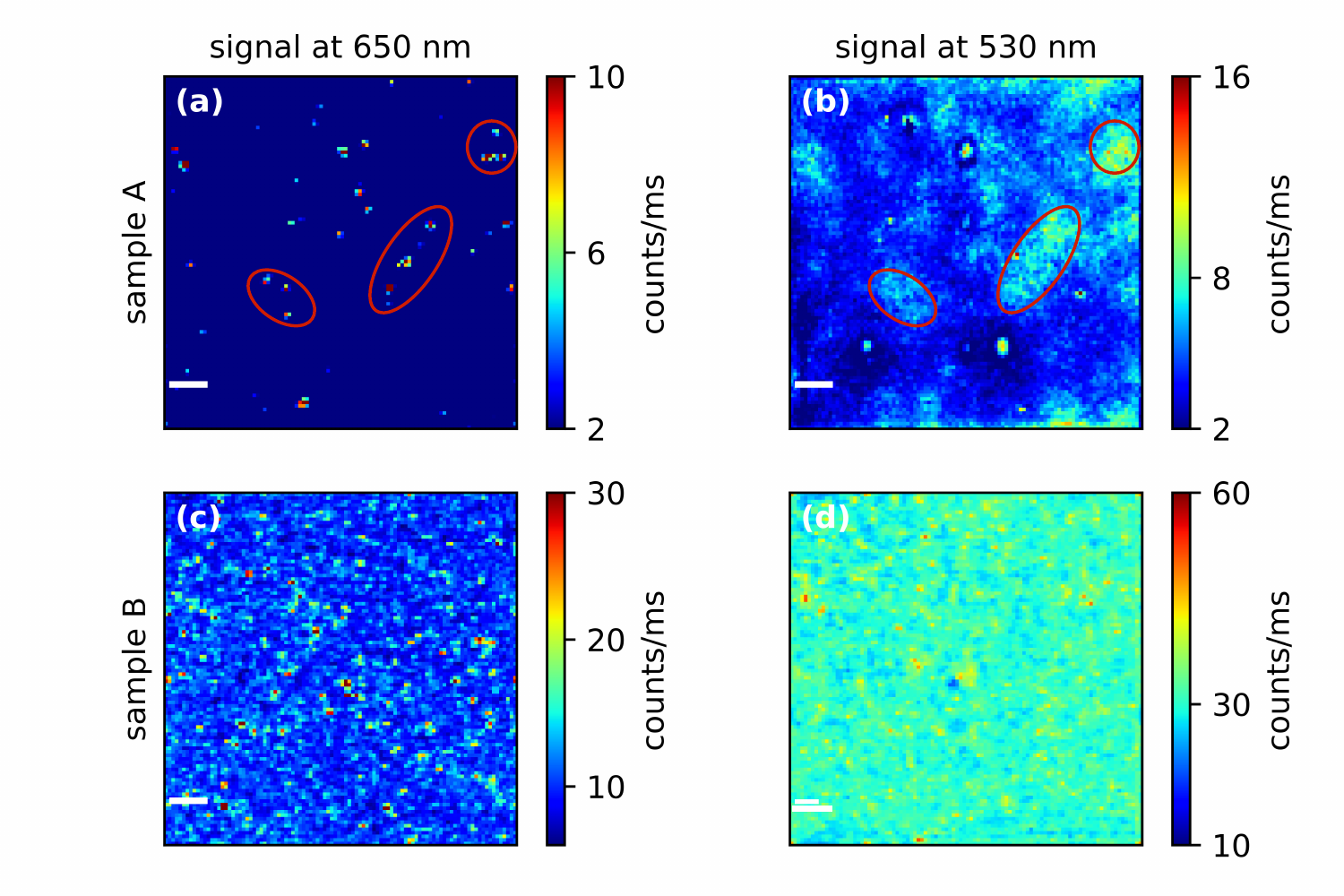}
\caption{(a) and (b): PL images of sample A recorded with a confocal microscope at the QD and \ce{MAPbBr3} emission wavelengths, respectively. Red ellipses indicate selected examples of the correlations between QD positions and an increased signal of the perovskite matrix. (c) and (d): PL images of sample B recorded with a confocal microscope at the QD and \ce{MAPbBr3} emission wavelengths respectively. The white scale bars represent 5\,\textmu m.}
\label{figure4}
\end{figure}

Figure~\ref{figure4} shows confocal photoluminescence images of a 50$\times$50\,\textmu m$^2$ region of sample A (Figs.~\ref{figure4}\,(a) and (b)) and sample B (Figs.~\ref{figure4}\,(c) and (d)); these images were recorded simultaneously at the \ce{MAPbBr3} and the QD emission wavelengths for both samples with pulsed-laser excitation (445\,nm, 2\,MHz repetition rate). Comparing the images obtained at the QD emission wavelength ($650\,\text{nm} \pm 15$\,nm, Figs.~\ref{figure4}\,(a) and (c), we find that sample A features small QD clusters (sub-micron size) with a rather low density, while sample B shows a more homogeneous distribution of the QDs with a larger signal. That latter property is coherent with the PLE signal of sample A and B presented in Fig.~\ref{figure2}\,(d) where the relative signal of the QDs excited at $\lambda = 445$\,nm is only about half for sample A as compared to sample B. We expect, however, from the nominal doping of the sample A (0.025\,\%) and sample B (0.1\,\%) a smaller QD density in sample A.
Focusing now on the signals recorded at the \ce{MAPbBr3} emission wavelength of ($530 \pm 5$)\,nm, Fig.~\ref{figure4}\,(b) and (d), we find a surprisingly correlated behavior between the perovskite and the QD emission: an homogeneous spatial distribution of the PL signal for sample B (Fig.~\ref{figure4}\,(d)), while the perovskite emission is structured for sample A (Fig.~\ref{figure4}\,(b)). It is rather straightforward to recognize that the \ce{MAPbBr3} emission signal in sample A increases at the locations of QD clusters with an almost one-to-one correlation with the QDs positions. Red ellipses in Fig.~\ref{figure4}\,(a) and (b) indicate a few examples of such correlations. The influence of the QDs on the emission of the \ce{MAPbBr3} matrix extends over a length scale of approximately 2\,\textmu m, while the size of the QD clusters lies in the sub-micron range. At the higher QD density of sample B, an interaction that extends over around 2\,\textmu m produces a homogeneous image at the \ce{MAPbBr3} emission wavelength. Our tentative interpretation of this behavior is that the QDs act as crystallization seeds for the \ce{MAPbBr3} matrix. Regions of better crystallinity exhibit higher emission QY and thus lead to a higher PL signal. If this hypothesis is correct, one would also expect an increase of the \ce{MAPbBr3} emission lifetime at the location where the crystallinity of the matrix is increased (leading a smaller amount of defects).

\subsection{QD-in-Perovskite Hybrid: Fluorescence-Lifetime Imaging Mi\-cros\-copy}

\begin{figure}[t]
\centering
\includegraphics[clip,width=15cm]{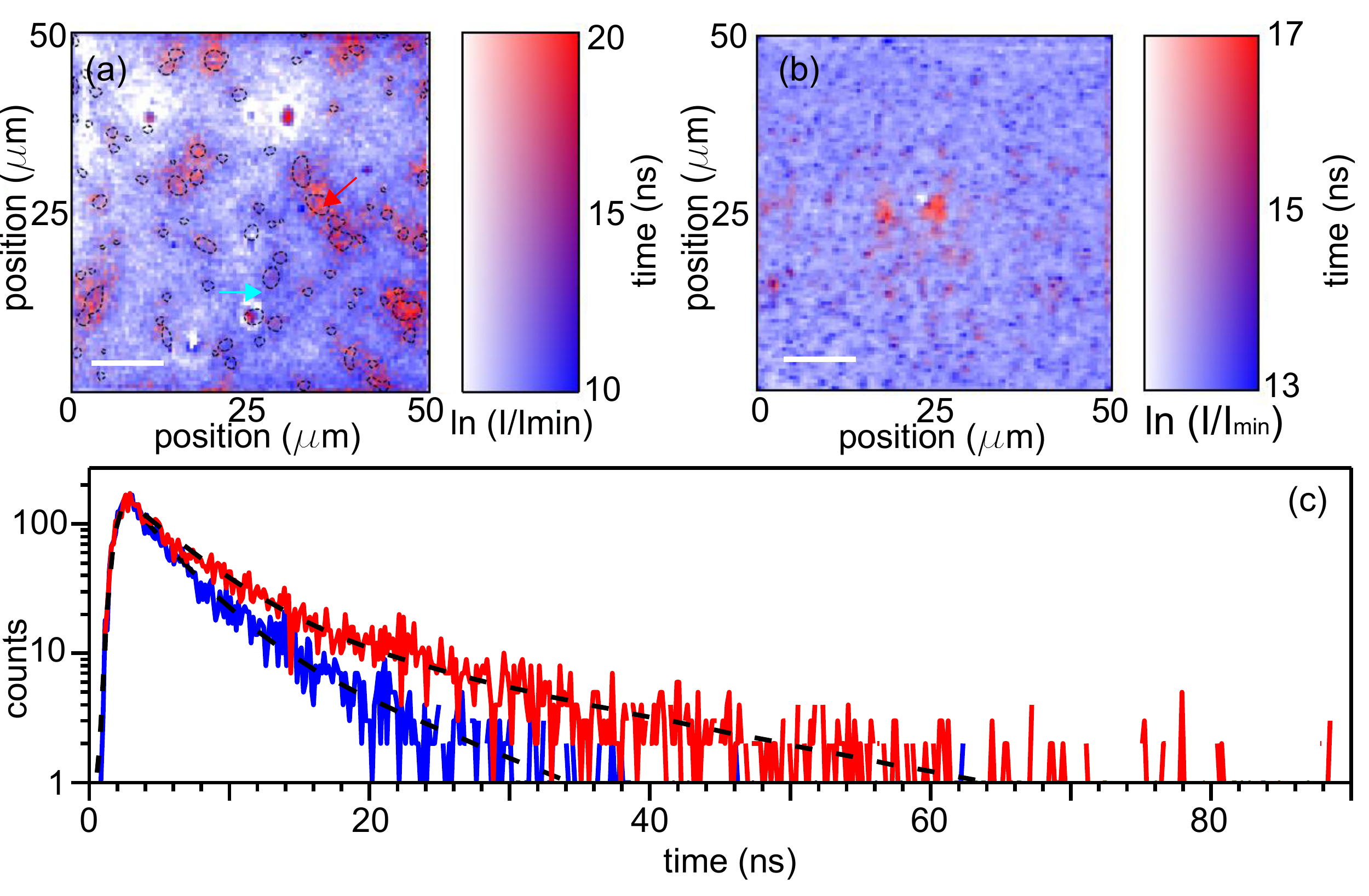}
\caption{(a) and (b): 50$\times$x50\,\textmu m$^2$ fluorescence-lifetime images of samples A and B, respectively, for the slow component of the \ce{MAPbBr3} emission wavelength. The positions of QD clusters are indicated by black-dashed circles in (a). (c): Two examples of lifetime curves and their corresponding bi-exponential fits (black dashed lines), extracted from points  $(x, y) = (35\,\text{\textmu m}, 26\,\text{\textmu m})$ (red curve) and (25\,\textmu m, 36\,\textmu m) (blue curve); the corresponding pixels are indicated by the two arrows in (a), the red arrow for the red curve, the light-blue arrow for the blue curve.}
\label{figure5}
\end{figure}

\begin{figure}[t]
\includegraphics[clip,width=12cm]{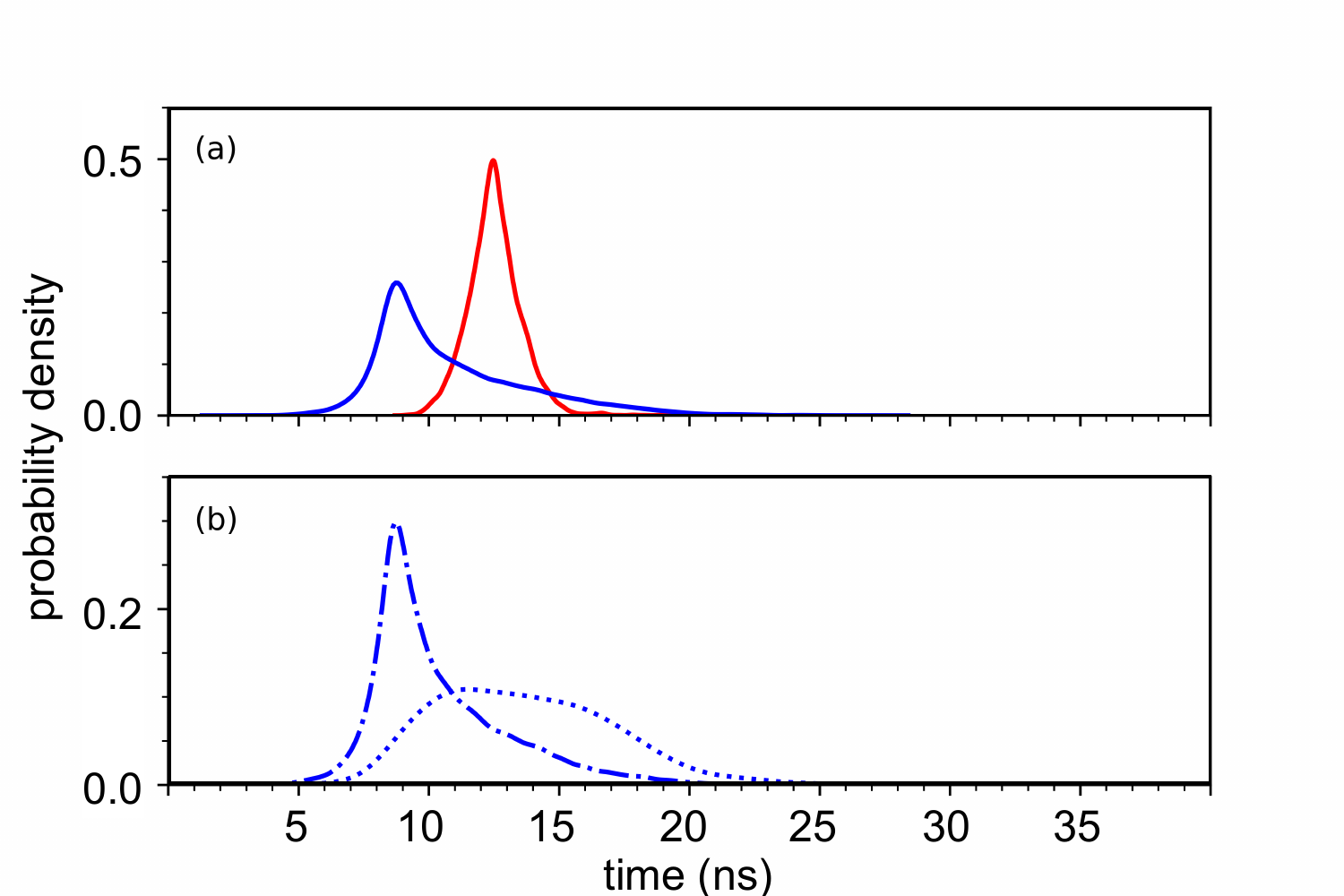}
\caption{(a): Distribution of the $\rm MAPbBr_3$ lifetimes extracted from the pixels in Fig.~\ref{figure5}\,(a), i.e. sample A (blue line) and Fig.~\ref{figure5}\,(b) i.e. sample B (red line). (b) Distribution of the \ce{MAPbBr3} lifetime extracted from the pixels in Fig.~\ref{figure5}\,(a) (sample A), post-selected by the strength of the QD luminescence at the same position, Fig.~\ref{figure4}\,(a), to distinguish high-intensity (at least 5\,counts/ms of QD signal, blue dotted line) and low intensity (less than 5\,counts/ms, blue dash-dotted line) pixels.}
\label{figure6}
\end{figure}

We present in Fig.~\ref{figure5} fluorescence-lifetime imaging measurements (FLIM) of samples A (Fig.~\ref{figure5}\,(a)) and B (Fig.~\ref{figure5}\,(b)) obtained at the \ce{MAPbBr3} emission wavelength, covering the same regions as shown as in Fig.~\ref{figure4}. Every pixel contains a lifetime curve which is fitted by a bi-exponential function with background, including convolution with the instrument response function (IRF):
\begin{equation}
I(t) = y_0 + A_1e^{-t/\tau_1} + A_2e^{-t/\tau_2}
\quad,
\end{equation}
%
%
%
%
where $y_0$ is the background, $\tau_1$ and $\tau_2$ the short and long lifetimes, respectively, and  $A_1$ and $A_2$ the associated amplitudes; the values of the slow component $\tau_1$ are represented in the FLIM images. The scale of these images enables an original presentation of the data with double color scaling: the saturation of the colors indicate the intensity level (white colors indicate minimum photon counts, while a saturated blue or red corresponds maximum of signal), while the colors themselves represent lifetime (blue is faster, red is slower). Fig.~\ref{figure5}\,(a) shows that the regions where the QDs are located (indicated by dashed circles) exhibit a longer lifetime (red pixels) of the \ce{MAPbBr3} emission as compared to QD-free zones (dominated by a blue color). Fig.~\ref{figure5}\,(c) presents two decay curves, one taken from the red pixel at the location indicated by the red arrow in Fig.~\ref{figure5}\,(a), the other from the blue pixel indicated by the light-blue arrow in Fig.~\ref{figure5}\,(a). The bi-exponential fits, taking into account the instrument response function, are shown as black dashed-lines in Fig.~\ref{figure5}\,(c) and yield a long component at $\tau^\text{red}_1$ = 15\,ns for the red pixel (close to a QD), and $\tau^\text{blue}_{1}$ = 9\,ns for the blue pixel (QD-free region). Fig.~\ref{figure5}\,(b) (sample B) shows a more homogeneous lifetime distribution than Fig.~\ref{figure5}\,(a), covering a smaller dynamical range and dominated by blue pixels. These results are compatible with our hypothesis that QDs locally modify the crystallinity of the \ce{MAPbBr3} matrix, provoking a local change in the emission lifetime. We present in Fig.~\ref{figure6}\,(a) the distribution of the lifetimes obtained from the pixels of Fig.~\ref{figure5}\,(a) (sample A, blue line) and (b) (sample B, red line). The lifetime distribution in sample A is highly skewed toward the long-lifetime tail, peaking at 8.5\,ns and extending to more than 20\,ns. On the contrary, the lifetime distribution in sample B is an almost symmetric super-Gaussian (leptokurtic) curve centered at 13\,ns with a FWHM of $\sim 2$\,ns. In order to unravel the lifetime distribution of sample A as a function of QD presence, we have separated the pixels of the lifetime image, Fig.~\ref{figure5}\,(a), based on an intensity-threshold criterion for the image of the QD luminescence of the same sample, see Fig.~\ref{figure4}\,(a). As can be seen in Fig.~\ref{figure6}\,(b), the distribution of lifetimes found for pixels exhibiting strong QD luminescence (dotted blue line) differs significantly from the one in regions of low QD signal (dash-dotted blue line): The short lifetimes making up the peak at 8.5\,ns are associated with the areas of the \ce{MAPbBr3} matrix that do not contain many QDs, while a broader distribution with a median value of 14\,ns corresponds to the QD-rich regions. We furthermore note that the median value of the lifetime distribution from QD-rich pixels in sample A corresponds to the maximum of the lifetime distribution found in sample B, which shows a stronger and spatially homogeneous QD signal. These results, showing an increased emission lifetime of the \ce{MAPbBr3} matrix at the QD locations, corroborate our speculative interpretation that the QDs act as local seeds of the \ce{MAPbBr3} host and help to locally improve crystallinity. 
\subsection{Single QDs in Perovskite}
\begin{figure}[t]
\includegraphics[clip,width=15cm]{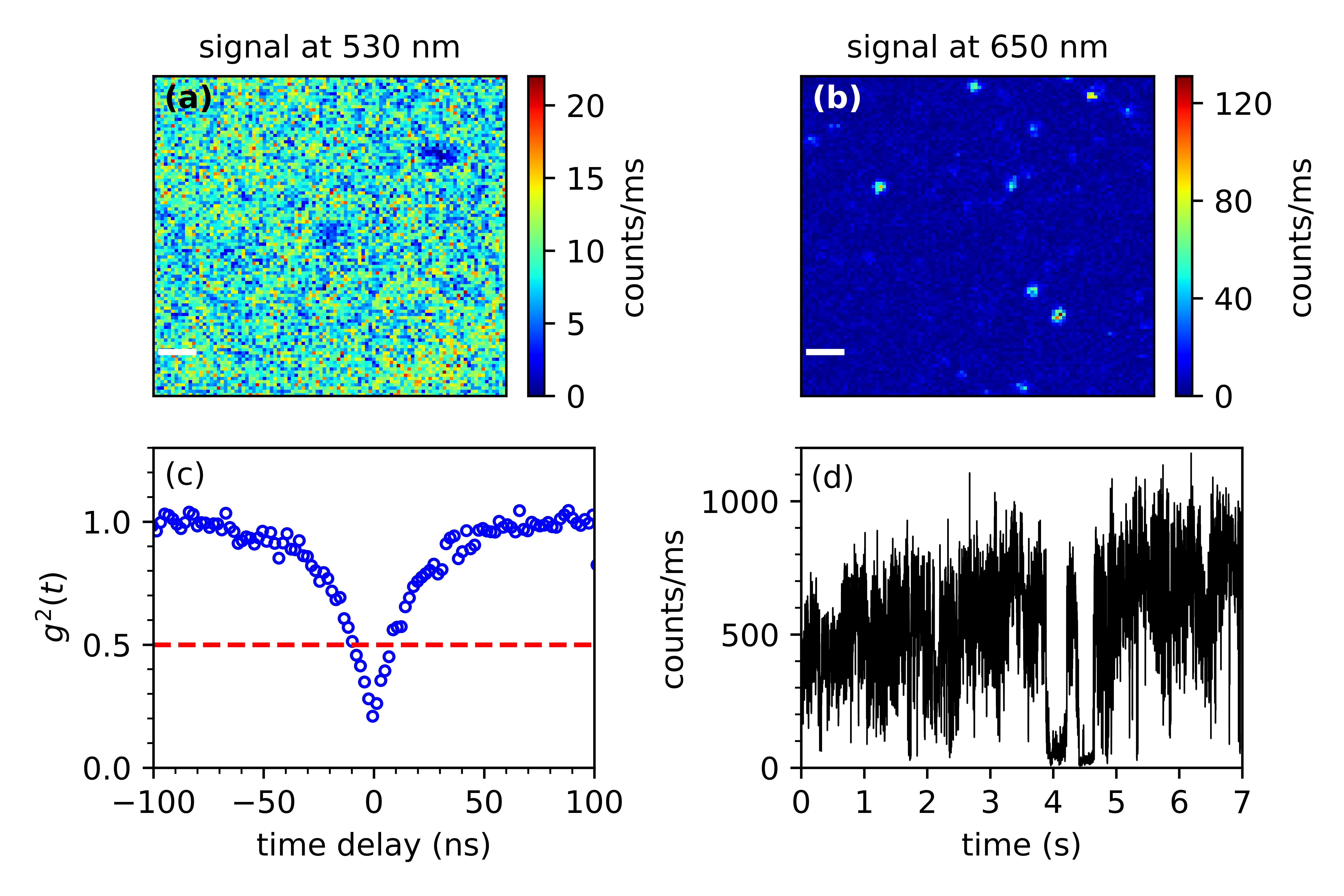}
\caption{(a) and (b): 10$\times$10\,textmu m$^2$ PL images recorded simultaneously at the \ce{MAPbBr3} (a) and QDs (b) emission wavelengths. The white scale bar represents 1\,\textmu m. (c): Antibunching of the luminescence of a single QD in perovskite. (d): Single QD-in-perovskite hybrid timetrace exhibiting blinking and a count rate of up to 800\,kHz.}
\label{figure7}
\end{figure}

Having in sight single photon sources applications\cite{lounis00oct27}, we present in Fig.~\ref{figure7}\,(a) and (b) confocal images of another region of sample A obtained at an excitation wavelength of 561\,nm and recorded simultaneously at the \ce{MAPbBr3} (a) and QD (b) emission wavelengths. The perovskite image shows a noisy unstructured signal over the 10$\times$10\,\textmu m$^2$ area (as expected when for excitation in its transparency region), while the QD image is structured with resolution-limited (200\,nm) high-emission spots. We performed photon-coincidence measurements at short times of the signal originating from these spots using a Hanbury Brown and Twiss (HBT) interferometer; a typical result can be seen in Fig.~\ref{figure7}\,(c). The $g^{(2)}(t)$ curve shows clear antibunching behavior, characteristic for the emission of an individual quantum system acting as a single-photon source. Measuring the signal of a single QD without difficulties (integration time in Fig.~\ref{figure7}\,(b) is 1 ms/pixel) is further testament to the conservation of the fluorescence quantum yield during the doping process: At high excitation intensity, we registered up to 800\,000 counts/s from a single emitter in the QD-in-perovskite hybrid, Fig.~\ref{figure7}\,(d). We performed a statistical analysis of 47 single QDs and fitted the obtained $g^{(2)}(t)$ functions with a single exponential rise functions\cite{lounis00oct27} :
\begin{equation}
g^{(2)}(t) = 1 - e^{-\vert t \vert / \tau_\text{\tiny QD}} 
\end{equation}
We extract an average rise time of the QD-in-perovskite hybrid of $\langle \tau_\text{\tiny QD} \rangle = (12 \pm 4)$\,ns. Details of the fitting procedure can be found in Ref.~\citenum{Baronnier21}. We compare  this average rise time to the antibunching obtained from 14 single CdSe/CdZnS QDs deposited on glass, for which the same analysis yields an average time of $\langle \tau^\text{ref}_\text{\tiny QD} \,\rangle = (39 \pm 14)$\,ns. Since we detect up to 800\,000 photons/s from a single QD in perovskite, we do not consider this reduced rise time as the consequence of the opening of new non-radiative recombination channels, i.e., a decrease of the emission quantum yield. Rather, we attribute it to the increase of the refractive index surrounding the QDs, from QDs in air deposited on a glass slide, to QDs embedded in a \ce{MAPbBr3} matrix. It has been demonstrated that CdSe/ZnS QDs can be used as probe of the local refractive index of their environment, with a sphere of sensitivity of radius $\sim 50 - 100$\,nm\cite{lebihan08sep, aubret16nanos82317}, and that their decay time in a \ce{SiO2} matrix follow those predicted by an effective medium model\cite{Aubret_2020}. 
We use the virtual cavity (VC) model which can be written as:
\begin{equation} \label{eq:vc1}
\gamma_\text{vac}^{} = \frac{\gamma_\text{ref}^{}} {\, \bar
n_\text{ref}^{} \,} \left( \frac{3} {\, \bar n_\text{ref}^2 + 2
\,} \right)^{\!\!2}
\qquad\qquad\mathrm{and}
\end{equation}
\begin{equation} \label{eq:vc2}
\gamma_\text{mat}^{} = \left( \frac{\, \bar n_\text{mat}^2 + 2
\,} {3} \right)^{\!\!2} \bar n_\text{mat}^{} \, \gamma_\text{vac}
\qquad,
\end{equation}
where $\gamma_\text{ref}^{}$, $\gamma_\text{mat}^{}$ and
$\gamma_\text{vac}^{}$ are the exciton recombination rates of CdSe/CdZnS QDs spincast
on a glass coverslip, in the \ce{MAPbBr3} matrix and in vacuum,
respectively, while $\bar n_\text{ref}^{}$ and $\bar n_\text{mat}^{}$ are the \emph{effective} refractive indices experienced by QDs on glass and in the \ce{MAPbBr3} matrix.
For the reference sample ($n_\text{glass}^{} = 1.52$) we find $\bar n_\text{ref}^{} = 1.22$, while the effective refractive index for the QD-in-perovskite heterostructure ($n_\text{glass}^{} =1.52$, $n_\text{\ce{MAPbBr3}}$ = 2.10, determined by ellipsometry at 650\,nm) is $\bar n_\text{mat}^{} = 1.84$, considering an homogeneous distribution of the QDs in the thin-film. Inserting $\bar n_\text{ref}^{}$ and $\gamma_\text{ref}^{} = 0.026$\,ns$^{-1}$ (from the measured $\tau_\text{ref}^{} = 39$\,ns) into the VC model of Eq.~(\ref{eq:vc1}), we obtain $\gamma_\text{vac}^{} = 0.016$\,ns$^{-1}$, which corresponds to an exciton recombination lifetime in vacuum of $\tau_\text{vac}^{} =
64$\,ns. The value of $\gamma_\text{vac}^{}$ thus identified leads to an expected radiative lifetime for the QD-in-perovskite hybrid of $\tau_\text{hybrid} = 11$\,ns when used in Eq.~(\ref{eq:vc2}), which corresponds to our experimental value. This estimate was obtained with a assumed radius of influence of $\rm R=80$\,nm, in accordance with our previous results on CdSe-based QDs\cite{aubret16aug31,aubret16nanos82317,Aubret_2020} ranging between 50 and 100\,nm. We therefore consider this feature as another proof that the QDs have been integrated to the perovskite matrix and that we are able to observe single QDs in perovskites as efficient single photons sources in the visible range.
Having identified single QD-in-perovskite hybrids, we studied their blinking properties, as compared to CdSe/CdZnS QDs on glass. The details of the blinking analysis technique, presented on the QDs with OA ligands, have been published elsewhere\cite{Baronnier21}. Blinking statistics were recorded under the same excitation conditions for both the single QD-in-perovskite hybrids and the OA-capped QDs: $\rm \lambda_{exc} = 561\,nm$ and $\rm P = 1\,\mu W$. We compare the results obtained on 11 single QD blinking from sample A and 24 single QDs with OA ligands deposited on glass. We find that the median proportion of ON times for QD-in-perovskite is as large as 99\,\%, and stands at 81\,\% for OA-capped reference QDs. This particularly large number for single QDs in perovskite is sample dependent, and we found it to be 94 and 96\,\%, respectively, on two other samples of QD-in-perovskites. This small variability of the results surely depends on a lot of factors, crystallisation being probably an important one; nevertheless, we found that encapsulated single QDs in hybrid perovskite show a consistent improvement of their blinking statistics with the median time spent in the ON state increasing in all three samples compared with the reference OA-capped QDs. Optimization of the blinking statistics by tuning deposition conditions will require further studies, but we identify the present results encouraging as a first step toward the control of blinking.


\section{CONCLUSION}

In conclusion, we have detailed a soft-chemistry method to realize the first doping of \ce{MAPbBr3} thin films with halide-capped (\ce{Cl} and \ce{Br}) CdSe/CdZnS QDs emitting in the visible. The QD shell needed a high ratio of Zn to avoid potential cation exchange between Cd and Pb ions of the matrix. A ligand-exchange procedure was developed, which removed the initial organic ligands (OA) and replaced them with halide ions, thus ensuring a direct and stable mixing of the QDs in DMF with the \ce{MAPbBr3} precursors. The obtained QD-in-perovskite films were $\sim$100\,nm thick on average, with cubic crystalline structure. Different CdSe/CdZnS doping levels were explored down to 0.025\,\% doping, ensemble PL spectra revealed the presence of emitting QDs at 638\,nm in the doped films. We focused our study on two samples (0.1 and 0.025\,\% doping level) processed under the same nominal conditions which produce comparable PL spectra, while their PLE was found to be substantially different. Their PLE spectra were explained with three additional relaxation processes, occurring in three wavelength regions. First, at high excitation energy, relaxation channels of the electrons from the QDs are open in the \ce{MAPbBr3} matrix; second, at an excitation energy around that of the \ce{MAPbBr3} exciton, this energy can be transferred to the QDs to increase their emission signal; and finally, in a narrow energy region where shallow defects of the \ce{MAPbBr3} matrix exist, electrons can be transferred from the QDs to those defects. Local PL measurements in confocal geometry show that the \ce{MAPbBr3} matrix emission signal increases at the QD locations in our doped thin-films, pointing toward a better crystallinity triggered the inclusion of the QDs, which could act as seeds in the \ce{MAPbBr3} crystallization process. The influence radius of the QD seeds on the \ce{MAPbBr3} is estimated to be $\pm \, 2$\,\textmu m. This hypothesis was further corroborated by FLIM measurements, which demonstrated that the emission lifetime of the \ce{MAPbBr3} matrix increased at the QD locations, pointing to a higher local quantum yield that can be explained by a better local crystallinity. We furthermore demonstrate that individual QDs in perovskite act as single-photon emitters and that their retained emission quantum yield enabled us to detect up to 800\,000 photons/s from a single hybrid, revealing new type of hybrid single-photon sources in the visible. Finally, we found that the single QD-in-perovskite hybrids have an enhanced blinking characteristics with up to 99\,$\%$ spent in the ON state, compared to 81\,\% for the reference QDs on glass. We believe this work can be a first step toward hybrid-based solar-cell concentrators with QDs emitting in the visible range (at high doping level), and a new way to envisage photonic devices for single-photon applications with colloidal QDs (low-doping level).

\paragraph{Acknowledgments.}
The authors thank J.-F. Sivignon, Y. Guillin, G. Montagne and the Lyon center for nano-opto technologies (NanOpTec) for technical support. This work was financially supported by the Agence Nationale de Recherche (ANR-16-CE24-0002), IDEXLYON of Université de Lyon in the framework ``investissement d'avenir'' (ANR-DEX-0005) and the Fédération de Recherche André Marie Ampère (FRAMA).




\providecommand{\latin}[1]{#1}
\makeatletter
\providecommand{\doi}
  {\begingroup\let\do\@makeother\dospecials
  \catcode`\{=1 \catcode`\}=2 \doi@aux}
\providecommand{\doi@aux}[1]{\endgroup\texttt{#1}}
\makeatother
\providecommand*\mcitethebibliography{\thebibliography}
\csname @ifundefined\endcsname{endmcitethebibliography}
  {\let\endmcitethebibliography\endthebibliography}{}


\end{document}